\documentclass[aps,pra,twocolumn,superscriptaddress,nofootinbib]{revtex4-1}

\usepackage{graphicx}
\usepackage{subfigure}
 
\usepackage{amsmath}
\DeclareGraphicsExtensions{.pdf,.jpeg,.png,.eps}

\usepackage{subfigure}
\usepackage{setspace}
 
\usepackage{array}
\usepackage{color}
\usepackage{rotating}
\usepackage{dcolumn}
 
\usepackage[colorlinks=true, allcolors=blue]{hyperref}
\usepackage{upgreek}
\usepackage[normalem]{ulem}

\renewcommand{\thetable}{\Roman{table}}

\newcolumntype{d}[1]{D{.}{\cdot}{#1} }

\definecolor{lightgray}{gray}{0.8}
\definecolor{llightgray}{gray}{0.95}

\def\0{^{(0)}}
\def\00{^{(0,0)}}
\def\##1{{\rm \bf #1}}
\def\=#1{\underline{\underline{#1}}}
\def\+#1{\underline{\bf #1}}
\def\*#1{\breve{\bf #1}}
\def\.{\mbox{ \tiny{$^\bullet$} }}

\def\le{\left(}
\def\ri{\right)}
\def\les{\left[}
\def\ris{\right]}
\def\lec{\left\{}
\def\ric{\right\}}

\def\Al2O3{\rm Al_{2}O_{3}}

\def\sfE{{\sf E}}

\def\eg{\sfE_{\rm g}}

\def\E0{E_{\scriptscriptstyle 0}}

\def\eg{\sfE_{\rm g}}

\def\Jdev{J_{\rm dev}}

\def\Jsc{J_{\rm sc}}

\def\LCIGS{L_{\rm CIGS}}	
\def\LCdS{L_{\rm CdS}}
\def\LZnO{L_{\rm ZnO}}
\def\LCZTSSe{L_{\rm CZTSSe}}

\def\LCdS{L_{\rm CdS}}
\def\LAZO{L_{\rm AZO}}

\def\LAl2O3{L_{\Al2O3}}
\def\LMgF2{L_{\rm \MgF2}}

\def\LMo{L_{\rm Mo}}

\def\MgF2{\rm MgF_2}

\def\N0{N_{\scriptscriptstyle 0}}

\def\nn1{n_{\rm 1}}

\def\N0{N_{\rm 0}}

\def\p1{p_{\rm 1}}

\def\Voc{V_{\rm oc}}

\def\Vext{V_{\rm ext}}

\def\Omegaph_{{\Omega_{\rm ph}}}
\def\Omegael_{{\Omega_{\rm el}}}

\def\egoCIGS{{\sf E}_{a1}}
\def\ACIGS{A_1}
\def\BCIGS{{\sf E}_{b1}}
\def\KCIGS{K_1}
\def\psiCIGS{\psi_1}
\def\alphaCIGS{\alpha_1}

\def\egoCZTSSe{{\sf E}_{a2}}
\def\ACZTSSe{A_2}
\def\BCZTSSe{{\sf E}_{b2}}
\def\KCZTSSe{K_2}
\def\psiCZTSSe{\psi_2}
\def\alphaCZTSSe{\alpha_2}

\def\CIGS{${\rm CuIn}_{1-{\xi_1}}{\rm Ga}_{\xi_1}{\rm Se}_2$}
\def\CZTSSe{${\rm Cu}_2{\rm ZnSn}\left({\rm S}_{\xi_2}{\rm Se}_{1-\xi_2}\right)_4$}

\def\FF{{\rm FF}}

\begin{document}

\title{Double-Absorber Thin-Film Solar Cell with 34\% efficiency}

\author{Faiz Ahmad}

\affiliation{Department of Engineering Science and Mechanics, Pennsylvania State University, University Park, PA 16802, USA}

\author{Akhlesh Lakhtakia}
\email[]{akhlesh@psu.edu}
\affiliation{Department of Engineering Science and Mechanics, Pennsylvania State University, University Park, PA 16802, USA}

\author{Peter B. Monk}
\affiliation{Department of Mathematical Sciences, University of Delaware, Newark, DE 19716, USA}

\date{\today}

\begin{abstract}
Power-conversion efficiency is a critical factor for the wider adoption of solar-cell modules. Thin-film solar cells are cheap and easy to manufacture, but their efficiencies are low compared to crystalline-silicon solar cells and need to be improved. A thin-film solar cell with two absorber layers (instead of only one), with bandgap energy graded in both, can capture solar photons in a wider spectral range. With a 300-nm-thick
    \CIGS~absorber layer and an 870-nm-thick  \CZTSSe~absorber layer, an
efficiency of $34.45\%$   is predicted by a detailed optoelectronic model, provided that the grading of bandgap energy is optimal in both absorber layers.

\end{abstract}
	

\maketitle

A photovoltaic solar cell consists of a metallic layer that serves as an optical reflector as well as
the electrical back-contact, a back-passivation layer, at least two semiconductor layers in which electrical charges are generated by the absorption of solar photons, an
electrical front-contact layer, and one or two antireflection coatings that are illuminated by the sun.
Some solar cells have an additional buffer or front-passivation layer as well, the role of any
passivation layer being to prevent recombination of two charges of opposite polarity.
When the solar cell is exposed to  sunlight,  photons with energy larger than the bandgap energy are absorbed in the semiconductor layers.  The absorbed energy excites electrons in the valence band. The excited electrons move to the conduction band and leave holes behind in the valence band. Thus, electron-hole pairs are created. When an electron and a hole recombine, energy is lost by conversion to heat and/or light. If a voltage is applied across the semiconductor layers, the electrons and the holes move in separate directions creating an electric current that depends
on the density of impurity atoms in each semiconductor layer. This is the basic principle of a photovoltaic solar cell~\cite{Nelson-book}.

Laudable technological and economic developments made on the commercially dominant crystalline-silicon (c-Si) solar cells 
have dramatically decreased investment costs~\cite{Green_J},
in line with what is needed to cope with climate emergency~\cite{Hawken}.
Solar parks   take  land that could otherwise be used for other purposes such as farming and  the transportation of electrical energy   adds transmission losses. In addition to a grid of solar parks,
 there is a need for in-device energy generation   of electricity for human progress to become truly unconstrained by energy economics.  
 
Thin-film solar cells can fill the need for in-device microwatt-scale ubiquitous generation of electricity.  However, the efficiencies of thin-film solar cells are lower than those of c-Si solar cells.   The highest reported efficiencies of thin-film CdTe, \CIGS (i.e., CIGS), and \CZTSSe (i.e., CZTSSe) solar cells are $21.0\%$, $22.6\%$, and $12.6\%$, respectively~\cite{Green2018}, while the efficiency of the c-Si solar cell is $26.7\%$~\cite{Green2018}. Thin-film solar cells require improvements.

In a typical solar cell, there is one thick semiconductor layer that is the dominant site for photons to be absorbed and is therefore the
major contributor to the electric current generated. The absorber
layer can be either an n-type or a p-type semiconductor. c-Si, CIGS, and CZTSSe solar cells have a single p-type thick absorber layer. The bandgap energy of the absorber
layer plays a crucial role in the   efficiency of the solar cell to convert solar (photonic) energy into electrical energy.  The  short-circuit current density $\Jsc$ is high/low but the 
open-circuit voltage $\Voc$ is low/high when the absorber layer has small/large bandgap energy~\cite{Green-PE2019}. Hence, the efficiency $\eta$ of a solar cell can be improved by the optimal design of the absorber layer. 

The bandgap energy of an absorber
layer made of a compound semiconductor such as CIGS and CZTSSe can be optimally
fixed within reasonable upper and lower bounds by correctly choosing the  composition of the semiconductor~\cite{Frisk2014, Woo2013, Schleussner2011, Hutchby1975}. Thus, the bandgap energy of CIGS depends on $\xi_1\in[0,1]$ and that of CZTSSe on $\xi_2\in[0,1]$.
Compositional grading of the absorber layer (i.e., grading $\xi_1$ for CIGS and $\xi_2$ for
CZTSSe) during fabrication can be exploited to grade the bandgap energy of that layer in the thickness direction. 
Linear  grading of the bandgap energy
of the absorber layer has been experimentally shown to improve the open-circuit voltage of CIGS solar cells~\cite{Frisk2014, Schleussner2011}. Similarly, it has been experimentally demonstrated that 
$\Jsc$ can be improved without reducing 
$\Voc$ by   grading the bandgap energy of the absorber layer in CZTSSe solar cells~\cite{Woo2013}. Detailed optoelectronic modeling
indicates that the proper grading of the bandgap energy of the absorber layer can enhance both 
$\Voc$ and $\Jsc$  in thin-film solar cells~\cite{Ahmad2019, Ahmad2020}; efficiencies as high as  $27.70\%$ and $21.74\%$ have been predicted
for CIGS solar cells~\cite{Ahmad2019}  and CZTSSe solar cells~\cite{Ahmad2020}, respectively.

Even with the bandgap-energy grading, the CIGS and CZTSSe absorber layers absorb only a part
of the optical energy available in the solar spectrum. The bandgap energy of CIGS  can
be varied between $0.947$~eV and $1.626$~eV, and that of CZTSSe between $0.91$~eV and $1.49$~eV.
One way to absorb  solar photons in a wider spectral range is to combine a CIGS solar cell and a CZTSSe solar cell in a tandem structure~\cite{Green-PE2019}. However, the current densities created in the two constituent solar cells will be different, and a two-terminal device  
with both solar cells in series will not be efficient~\cite{Green-PE2019}. Combining a
CIGS solar cell and a CZTSSe solar cells in a four-terminal device  will
require additional circuitry to be fabricated and managed, resulting in parasitic losses, effectively reducing the overall efficiency of the tandem solar cell. 
We propose here another option to harvest  photons over a wider spectral range and improve the  performance of   thin-film solar cells.

\begin{figure}[h]
	\centering
	\includegraphics[width=0.9\columnwidth]{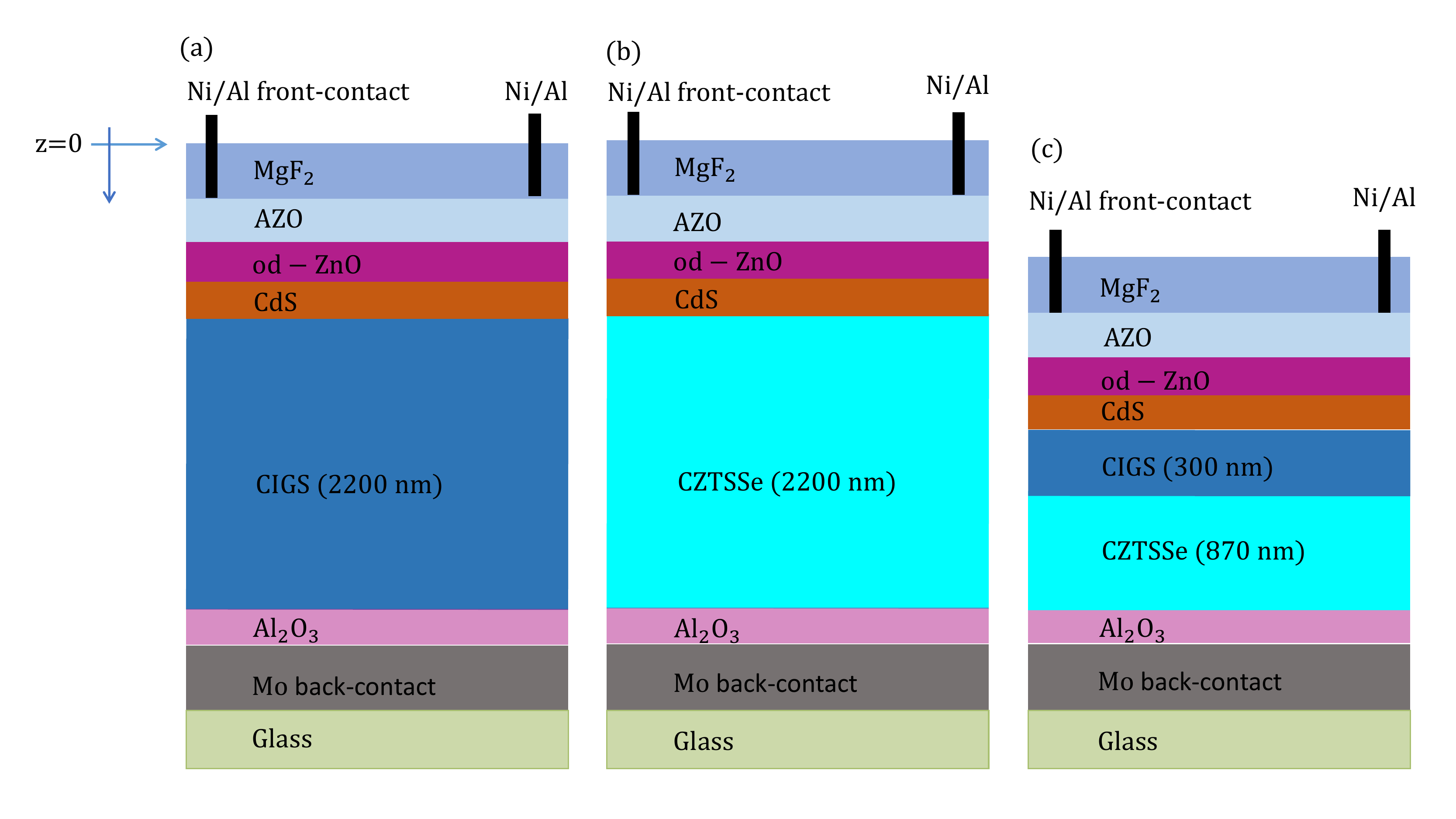}    
	\caption{Schematic of the thin-film solar cell based on (a) CIGS absorber layer (b) CZTSSe absorber layer (c) CIGS-CZTSSe double-absorber layer.  The thickness of the absorber layer in  both (a) and (b) equals 2200~nm for the highest efficiency reported in the literature~\cite{Green2018}. The optimal  thicknesses of the two absorber layers in (c) are predicted in this paper.
		\label{figure1}}
\end{figure}

The structures of the CIGS and CZTSSe solar cells---shown in Figs.~\ref{figure1}(a)
and ~\ref{figure1}(b), respectively----are identical, except for having a p-type absorber layer
made of either CIGS or CZTSSe. Solar cells of both types
have a molybdenum (Mo) back-contact layer,
an aluminum-oxide ($\Al2O3$) back-passivation layer, the absorber layer of a p-type
semiconductor, a semiconductor layer of n-type cadmium sulfide (CdS), an oxygen-deficient
zinc-oxide (od-ZnO) front-passivation layer, a front-contact layer made of aluminum-doped
zinc oxide (AZO), and an antireflection coating of magnesium fluoride ($\MgF2$).

If absorber layers of both types were
present in a single two-terminal solar cell, parasitic impedances and
 additional circuitry will be avoided. However, the absorbers of both
 types must have minimum lattice difference and should be capable of being deposited in a single device with compatible fabrication techniques. CIGS and CZTSSe  are almost lattice matched~\cite{Adachi-book2015, Klinkert2014} and can be fabricated using  vapor deposition techniques~\cite{Chopra2004}. 
 
Hence, the double-absorber CIGS-CZTSSe solar cell shown in Fig.~\ref{figure1}(c) is proposed and theoretically studied in this
communication, using a detailed optoelectronic model \cite{Anderson-JCP2020}
coupled with the differential evolution algorithm \cite{Storn1997}
for optimization. In conformance with existing thin-film solar cells, the thicknesses of various layers were
fixed as follows: $L_{\MgF2}=110$~nm, $\LAZO=100$~nm, 
$\LZnO=80$~nm, $\LCdS =70$~nm, $\LAl2O3=10$~nm, and $\LMo =500$~nm. Also in conformance with existing solar cells, the thicknesses
$\LCIGS\leq2200$~nm and $\LCZTSSe\leq2200$~nm  of the
two absorber layers were kept variable. With the exposed surface of
the $\MgF2$ layer identified as the plane $z=0$ and the $z$ axis pointing into the solar cell
(as shown in Fig.~\ref{figure1}), the $z$-dependent bandgap energy (in eV)
was modeled in the CIGS layer  as \cite{Ahmad2019}
\begin{align} 
	\nonumber
	&\eg(z)=\egoCIGS +\ACIGS\le \BCIGS-\egoCIGS\ri \, 
	\\
	\nonumber
	&\times
	\le \frac{1}{2}\, \lec \sin\les 2\pi \le\KCIGS \frac{z-L_1}{\LCIGS}-\psiCIGS\ri \ris+1\ric\, \ri^{\alphaCIGS} \, , 
	\\[5pt]
	&\qquad\qquad  z\in\left[L_1,L_2\right]\, ,
	\label{Eqn:Sin-bandgap-CIGS}
\end{align}
and in the CZTSSe layer as \cite{Ahmad2020}
\begin{align} 
	\nonumber
	&\eg(z)=\egoCZTSSe +\ACZTSSe\le \BCZTSSe-\egoCZTSSe\ri \, 
	\\
	\nonumber
	&\times
	\le \frac{1}{2}\, \lec \sin\les 2\pi \le\KCZTSSe \frac{z-L_2}{\LCZTSSe}- \psiCZTSSe\ri\ris +1\ric\, \ri^{\alphaCZTSSe} \, , 
	\\[5pt]
	&\qquad\qquad  z\in\left[L_2,L_3\right]\, ,
	\label{Eqn:Sin-bandgap-CZTSSe}
\end{align}
where
$\BCIGS=1.626$~eV, $\BCZTSSe=1.49$~eV,
$L_1=L_{\MgF2}+\LAZO+\LZnO+\LCdS$,
$L_2=L_1+\LCIGS$, and
$L_3=L_2+\LCZTSSe$. Whereas $\eg(z)$ in the CIGS layer can be engineered through
$\xi_1(z)$ \cite{Frisk2014}, $\eg(z)$ in the CZTSSe layer can be engineered through $\xi_2(z)$
\cite{Adachi-book2015,Kanevce}.

Equations (\ref{Eqn:Sin-bandgap-CIGS}) and 
(\ref{Eqn:Sin-bandgap-CZTSSe}) can represent a wide variety of bandgap-energy
profiles. Based on experience, optimization was carried out in the parameter space defined
as follows: $\egoCIGS\in[0.947,1.626]$~eV, 
$\ACIGS\in[0,1]$,
$\KCIGS\in[0,8]$, $\psiCIGS\in[0,1]$, $\alphaCIGS\in[0,7]$,
$\egoCZTSSe\in[0.91,1.49]$~eV, 
$\ACZTSSe\in[0,1]$,
$\KCZTSSe\in[0,8]$, $\psiCZTSSe\in[0,1]$, and $\alphaCZTSSe\in[0,8]$,
$\LCIGS\in[0,2200]$~nm, $\LCZTSSe\in[0,2200]$~nm, and $0<\LCIGS+\LCZTSSe\leq2200$~nm.

The  optoelectronic model \cite{Anderson-JCP2020}
has a photonic step and an electronic step. In the photonic step, the transfer-matrix method~\cite{Berreman1972,TEMM2020-book}  was used to determine the electric and magnetic fields everywhere inside the solar cell due to  normally  incident   monochromatic radiation. The transfer-matrix method is an efficient 
 technique to solve the frequency-domain Maxwell equations.   Thereafter, the electron-hole-pair generation rate $G(z)$ was determined in the   ZnO, CdS, CIGS, and CZTSSe layers   of the double-absorber solar cell~\cite{Ahmad2019}, assuming normal illumination by unpolarized polychromatic light endowed with the AM1.5G solar spectrum~\cite{SSAM15G}. The frequency-dependent relative permittivity of every material in the double-absorber solar cell is available elsewhere \cite{Ahmad2019,Ahmad2020}.
 
In the electronic step, the electron-hole-pair generation rate was used as an input to the 1D 
drift-diffusion equations~\cite{Nelson-book, Anderson-JCP2020} applied to the semiconductor layers. 
The nonlinear Shockley--Read--Hall, Auger, and radiative
contributions to the electron-hole recombination rate $R(z)$ were incorporated
for the ZnO, CdS, CIGS, and CZTSSe layers. Both electrical contacts were assumed to be ideally ohmic and  local quasi-thermal equilibrium was applied to determine boundary conditions.
Electrical data for  ZnO, CdS, CIGS, and CZTSSe  are available elsewhere \cite{Ahmad2019,Ahmad2020}.
A set of six nonlinear differential equations
was solved using a hybridizable discontinuous Galerkin (HDG) scheme~\cite{Anderson-JCP2020, Brinkman2013, Brezzi2002} to determine the  current density $\Jdev$ and the electrical power density $P=\Jdev\Vext$ as functions of the bias voltage $\Vext$ under steady-state
conditions. In turn, the $\Jdev$-$\Vext$ and the $P$-$\Vext$ curves yielded $\Jsc$, $\Voc$,  $\eta$,
and a figure of merit called the fill factor $\FF\in[0,1]$ which should be as high as possible~\cite{Nelson-book}. The model has been validated against experimental results \cite{Ahmad2019,Ahmad2020}.
 
Finally, the widely used differential evolution algorithm~\cite{Storn1997} was adopted to maximize $\eta$ with respect to $\egoCIGS$, 
$\ACIGS$,
$\KCIGS$, $\psiCIGS$, $\alphaCIGS$,
$\egoCZTSSe$,
$\ACZTSSe$,
$\KCZTSSe$, $\psiCZTSSe$,  $\alphaCZTSSe$,
$\LCIGS$, and $\LCZTSSe$. The algorithm was implemented using
MATLAB\textsuperscript{\textregistered} version R2019a and run over a  search time of eight weeks. 
Given an initial guess in this
search space, the underlying strategy in differential evolution is to improve
the candidate solution at every iteration step and
does not require explicit gradients of the cost
function (i.e., $\eta$). A multidimensional parameter space  
can be searched by this  metaheuristic
algorithm.

The highest value of $\eta$ predicted for the double-absorber solar cell is $34.45\%$; correspondingly,
 $\Jsc=38.11$~mA~cm$^{-2}$, $\Voc=1085$~mV, and  $\FF=0.83$.
The optimal thicknesses of the absorber layers in the double-absorber solar cell are $\LCIGS=300$ nm and $\LCZTSSe =870$ nm. The corresponding bandgap-energy parameters are as follows:
 $\egoCIGS=0.95$~eV, 
$\ACIGS=0.91$,
$\KCIGS=1.88$, $\psiCIGS=\psiCZTSSe=0.75$, $\alphaCIGS=\alphaCZTSSe=6$,
$\egoCZTSSe=0.91$~eV,
$\ACZTSSe=0.99$, and
$\KCZTSSe=2$.  
The efficiency drops to no less than $34.43\%$, if any of the optimal bandgap-energy parameters is altered by $1\%$.

 If the CZTSSe absorber layer is absent but $\LCIGS=300$~nm, the highest efficiency predicted 
  is $19.01\%$, $\Jsc=25.98$~mA~cm$^{-2}$,  $\Voc=1023$~mV, and $\FF=0.73$; these values
  were obtained with $\egoCIGS=0.95$~eV, 
$\ACIGS=0.98$,
$\KCIGS=1.5$, $\psiCIGS=0.74$, and $\alphaCIGS=6$.
If the CIGS absorber layer is absent but $\LCZTSSe=870$~nm, the highest efficiency is $21.74\%$, $\Jsc=37.39$~mA~cm$^{-2}$,   $\Voc=772$~mV, and  $\FF=0.75$; these values
  were obtained with $\egoCZTSSe=0.92$~eV, 
$\ACZTSSe=0.98$,
$\KCZTSSe=2$, $\psiCZTSSe=0.75$, and $\alphaCZTSSe=6$.
Notice that the double-absorber  solar cell outperforms both single-absorber cells in all four perfomance parameters: $\eta$, $\Jsc$,
$\Voc$, and $\FF$. The double-absorber solar cell appears to derive
the high value of $\Jsc$ from the CZTSSe absorber layer
and the high value of $\Voc$ from the CIGS absorber layer.

Equations (\ref{Eqn:Sin-bandgap-CIGS}) and 
(\ref{Eqn:Sin-bandgap-CZTSSe}) encompass both absorber layers being homogeneous
in the space of the parameters chosen for optimization. Therefore, the highest efficiency
with the graded-bandgap-energy absorber layers will necessarily exceed (or equal)
the highest efficiency with homogeneous-bandgap-energy absorber layers. Indeed,
if we fix $\LCIGS=300$~nm, $\LCZTSSe=870$~nm, and $\ACIGS=\ACZTSSe=0$,
the highest efficiency predicted is 
$11.08\%$ with $\egoCIGS=0.95$~eV and $\egoCZTSSe=0.91$~eV; correspondingly, $\Jsc=35.90$~mA~cm$^{-2}$,  $\Voc=455$~mV, and  $\FF=0.67$.  
If $\LCIGS=300$~nm and $\LCZTSSe=0$, 
 the highest efficiency predicted is $11.59\%$ with $\egoCIGS=1.25$~eV, $\Jsc=22.56$~mA~cm$^{-2}$, $\Voc=681$~mV,
 and $\FF=0.76$. Conversely, if   $\LCIGS=0$  and $\LCZTSSe=870$~nm, 
  the highest efficiency predicted is $11.84\%$ with 
$\egoCZTSSe=1.20$~eV, $\Jsc=30.13$~mA~cm$^{-2}$, $\Voc=558$~mV,
 and $\FF=0.70$. Thus, even though the double-absorber solar cell exceeds both single-absorber
 solar cells in $\Jsc$, it underperforms both in $\Voc$ so much so that its efficiency is somewhat lower than either's. Grading of the bandgap energy of both absorber layers is the key to significantly higher efficiency.

\begin{figure}[h] 
	\centering   
	\includegraphics[width=0.6\columnwidth]{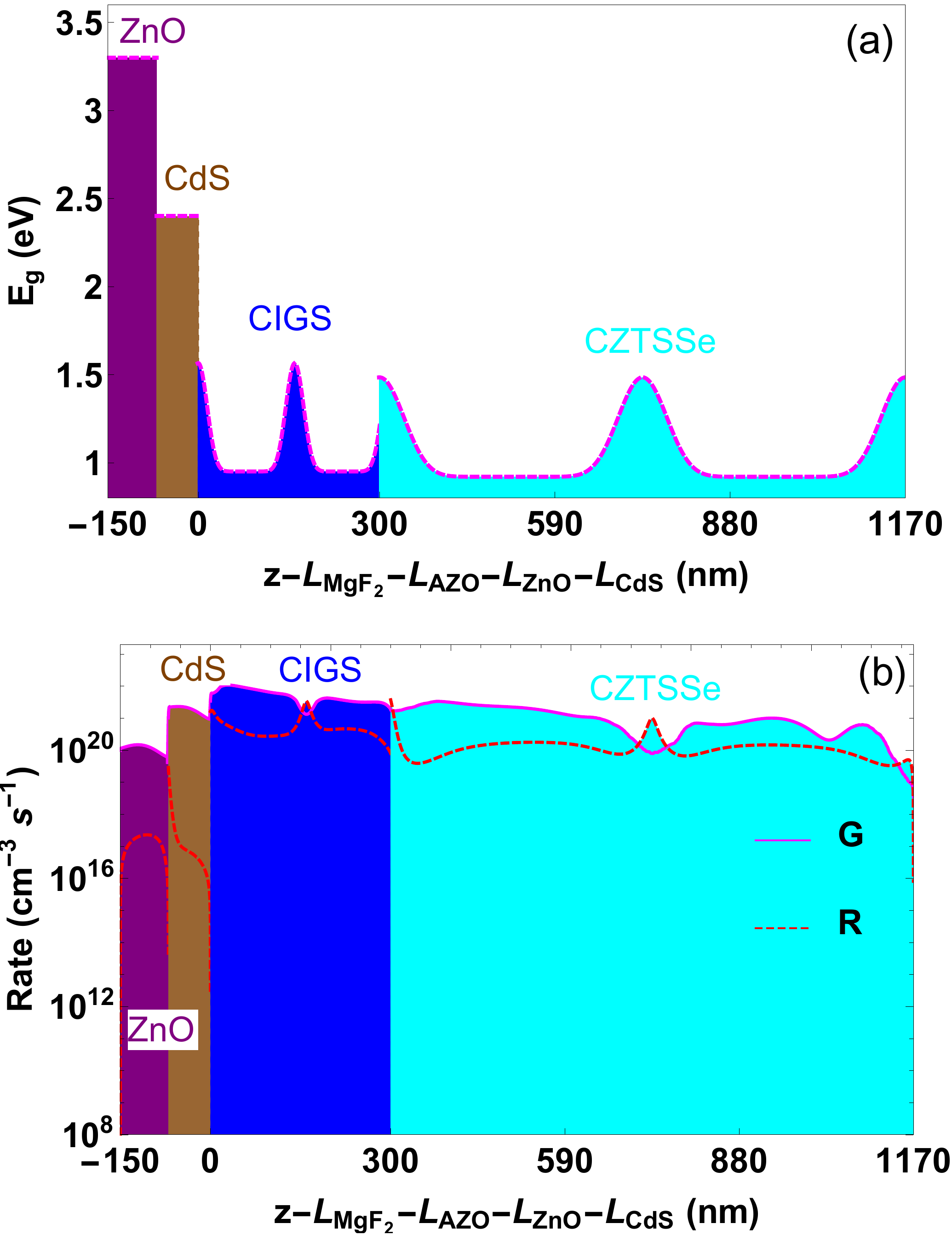}
	\caption{(a) $\eg(z)$ and (b) $G(z)$ and
	$R(z)$ in the ZnO/CdS/CIGS/CZTSSe region of the optimal double-absorber solar cell.
		\label{figure2}	 }
\end{figure}

The spatial variation of $\eg$ in the ZnO/CdS/CIGS/CZTSSe region of the optimal double-absorber solar cell is depicted in Fig.~\ref{figure2}(a), and those of $G(z)$ and $R(z)$
 in Fig.~\ref{figure2}(b). The generation rate is higher in regions with lower bandgap energy and \textit{vice versa}. The recombination rate is
higher in regions with higher bandgap energy, due to higher defect/trap density caused by higher gallium or sulfur content in those regions.

 Whereas $\eg$ is independent of $z$ in both the
ZnO and CdS layers (by design), it varies with $z$ in both absorber layers. This variation comprises constant-$\eg$ regions separated by regions with large $\eg$ gradients. The bandgap energy is low in the constant-$\eg$ regions, these regions being responsible for elevating the 
electron-hole pair generation rate because less energy is required
to excite an electron-hole pair across a narrower bandgap~\cite{Fonash-book}. Figure~\ref{figure2}(b)
confirms that $G(z)$ exceeds $R(z)$ in the constant-$\eg$ regions.

The large  $\eg$ gradient
close to the back surface in the CZTSSe layer enhances the drift field to reduce
the back-surface recombination rate, thereby supplementing the role
of the $\Al2O3$ passivation layer~\cite{Dullweber2001, Casper2016}. Since the bandgap energy is high close to both faces of  each absorber layer, $\Voc$ is high in the optimal design~\cite{Dullweber2001, Yang2016, Gloeckler-Sites2005JPCS}. The triangular regions in the
middle of each absorber layer in Fig.~\ref{figure2}(a)
 also create an additional drift  field that favors the charge-carrier collection deep inside the absorber layer~\cite{Hutchby1975}.

\begin{figure}[htb] 
	\centering   
	\includegraphics[width=0.52\columnwidth]{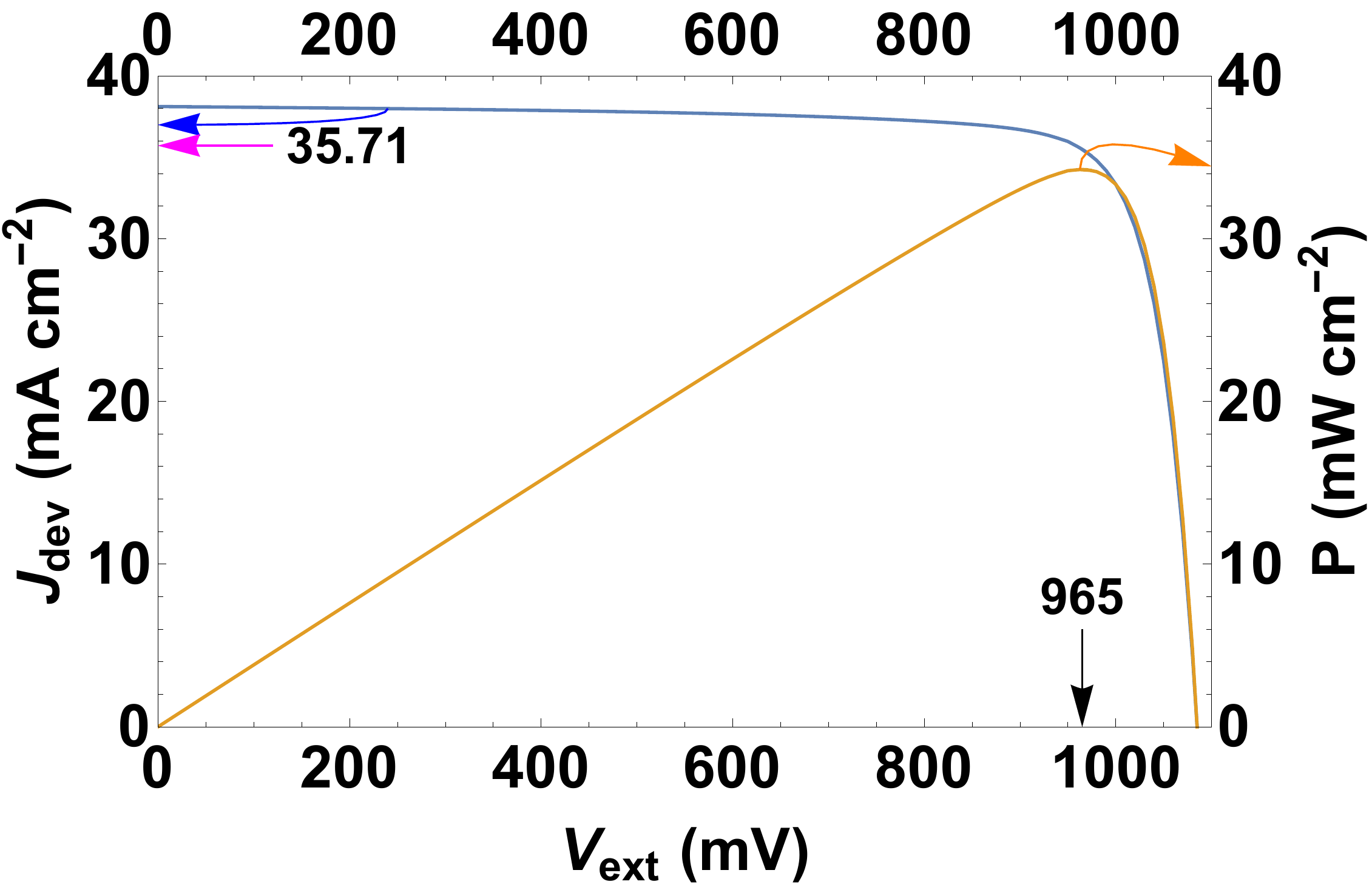} 
	\caption{Plots of $\Jdev$ and $P$ vs. $\Vext$ of the optimal double-absorber solar cell. 
The  values of $\Jdev$ and $\Vext$ for maximum $P$ 
		are  identified.
		\label{figure3}	 }
\end{figure}

The $\Jdev$-$\Vext$ characteristics of the optimal double-absorber solar cell are shown in Fig.~\ref{figure3}. The optoelectronic model predicts that the solar cell should be operated
with $\Vext=965$~mV to deliver
$\Jdev=35.71$~mA~cm$^{-2}$; then $34.45$~mW~cm$^{-2}$ is predicted as the maximum extractable power density when the incident solar flux is $100$~mW~cm$^{-2}$. 

To conclude, CIGS solar cells deliver higher efficiency than CZTSSe solar cells,
whether the bandgap energy in the absorber layer is homogeneous \cite{Green2018}
or graded \cite{Ahmad2019,Ahmad2020}. But CIGS contains indium which is not plentiful
on  our planet whereas no constituent element of CZTSSe  suffers from that
constraint. If absorber layers of both CIGS and CZTSSe are used in a single solar cell,
then the efficiency can be   boosted highly to $34.45\%$ and the fill factor to $0.83$ provided
the bangap energy is optimally graded,
with the CIGS   layer being only 300-nm thick and the CZTSSe layer being 870-nm
thick. Practical realization of this double-absorber thin-film solar cell, or an approximative
variant thereof, will require the attention of
experimentalists and may not perform as well as predicted
by a detailed optoelectronic model. Nevertheless, this  solar cell is promising for ubiquitous
in-device microwatt-scale  generation of electricity.

\noindent\textbf{Acknowledgments.} 
This research was  supported by: (i)  the US National 
Science Foundation  under grants numbered DMS-1619901 and DMS-1619904
and (ii) the Charles Godfrey Binder 
Endowment at Penn State.

The data that support the findings of this study are available within the article


\begin{thebibliography}{99}

\bibitem{Nelson-book}   
J.~Nelson,  \textit{The Physics of Solar Cells} (Imperial College Press, London, 
UK, 2003).
	

\bibitem{Green_J}
M. A. Green, \href{https://doi.org/10.1016/j.joule.2019.02.010}{\textit{Joule}}~\textbf{3}, 631  (2019).

\bibitem{Hawken}
P. Hawken (ed.), 
\textit{Drawdown}
(Penguin, New York, NY, USA, 2017).

\bibitem{Green2018} 
M. A. Green, E. D. Dunlop, J. Hohl-Ebinger, M. Yoshita, N. Kopidakis, and A. W.~Y. Ho-Baillie, \href{https://doi.org/10.1002/pip.3228}{\textit{Prog. Photovolt.: Res. Appl.}}~\textbf{28}, 3  (2020).

\bibitem{Green-PE2019} 
M.~A.~Green, \href{https://doi.org/10.1088/2516-1083/ab0fa8}{\textit{Prog. Energy}}~\textbf{1}, 013001 (2019).

\bibitem{Frisk2014}
C.~Frisk, C.~Platzer-Bj\"{o}rkman, J.~Olsson, P.~Szaniawski, J.~T. W\"{a}tjen, V.~Fj\"{a}llstr\"{o}m, P.~Salom\'e, and M.~Edoff,   \href{https://doi.org/10.1088/0022-3727/47/48/485104}{\textit{J. Phys. D: Appl. Phys.}}~\textbf{47}, 485104 (2014).	

\bibitem{Woo2013} 
K.~Woo, Y. Kim, W.~Yang, K.~Kim, I. Kim, Y.~Oh, J.~K.~Kim, and J.~Moon, \href{https://doi.org/10.1038/srep03069 }{\textit{Sci. Rep.}}~\textbf{3}, 03069 (2013).

\bibitem{Hutchby1975}
J.~A.~Hutchby, \href{https://doi.org/10.1063/1.88208}{\textit{Appl. Phys. Lett.}}~\textbf{26}, 457  (1975).

\bibitem{Schleussner2011}
S. Schleussner, U. Zimmermann, T. W\"{a}tjen, K. Leifer, and M. Edoff, \href{https://doi.org/10.1016/j.solmat.2010.10.011}{\textit{Sol. Energy Mater. Sol. Cells}}~\textbf{95}, 721  (2011).

\bibitem{Ahmad2019}  
F.~Ahmad, T.~H.~Anderson, P.~B.~Monk, and A.~Lakhtakia, \href{https://doi.org/10.1364/ao.389988}{\textit{Appl. Opt.}} {\bf58}, 6067  (2019); erratum: {\bf 59}, 2615 (2020).

\bibitem{Ahmad2020} 
F. Ahmad, A. Lakhtakia, T. H. Anderson, and P. B. Monk, \href{https://doi.org/10.1088/2515-7655/ab6f4a }{\textit{J. Phys.: Energy}}~\textbf{2}, 025004 (2020).

\bibitem{Adachi-book2015} 
S. Adachi,  \textit{Earth-Abundant Materials for Solar Cells} (Wiley, Chichester, West Sussex, UK, 2015).	

\bibitem{Klinkert2014} 
T. Klinkert, M. Jubault, F. Donsanti, D. Lincot, and J.-F. Guillemoles, \href{https://doi.org/10.1016/j.tsf.2014.02.071}{\textit{Thin Solid Films}} \textbf{558}, 47  (2014). 

\bibitem{Chopra2004} 
K. L. Chopra, P. D. Paulson, and V. Dutta, \href{https://doi.org/10.1002/pip.541}{\textit{Prog. Photovolt.: Res. Appl.}}~\textbf{12}, 69  (2004).

\bibitem{Anderson-JCP2020}
T. H. Anderson, B. J. Civiletti, P. B. Monk, and A. Lakhtakia, \href{https://doi.org/10.1016/j.jcp.2020.109242}{\textit{J. Comput. Phys.}}~\textbf{407},  109242 (2020).

\bibitem{Storn1997}
R.~Storn and K.~Price, \href{https://doi.org/10.1023/A:1008202821328}{\textit{J. Global Optim.}}~\textbf{11}, 341  (1997).


\bibitem{Kanevce}
A. Kanevce, I. Repins, and S.-H. Wei,
\href{http://dx.doi.org/10.1016/j.solmat.2014.10.042}{\textit{Sol. Energy Mater. Sol. Cells}} \textbf{133}, 119  (2015).

\bibitem{Berreman1972}
D. W. Berreman,  \href{https://doi.org/10.1364/josa.62.000502}{\textit{J. Opt. Soc. Am.}} \textbf{62}, 502  (1972).

\bibitem{TEMM2020-book}
T. G. Mackay and A. Lakhtakia, \textit{The Transfer-Matrix Method in Electromagnetics and Optics} (Morgan \& Claypool, San Rafael, CA, USA, 2020).



\bibitem{SSAM15G}
NREL, \href{http://rredc.nrel.gov/solar/spectra/am1.5/}{Reference Solar Spectral Irradiance: Air Mass 1.5} (accessed 09 June 2020).


\bibitem{Brinkman2013}
D.~Brinkman, K.~Fellner, P.~Markowich, and M.-T. Wolfram, \href{https://doi.org/10.1142/S0218202512500625}{\textit{Math. Models Methods Appl. Sci.}} {\bf 23}, 839  (2013).

\bibitem{Brezzi2002}
F.~Brezzi, L.~D. Marini, S.~Micheletti, P.~Pietra, R.~Sacco, and S.~Wang, 
\href{https://doi.org/10.1016/S1570-8659(04)13004-4}
{\textit{Handbook of Numerical Analysis}} \textbf{13},  317  (2005).

\bibitem{Fonash-book}
S. J. Fonash, \textit{Solar Cell Device Physics}, 2nd ed., (Academic, Burlington, MA, USA, 2010).

\bibitem{Dullweber2001}T.~ Dullweber, O.~Lundberg, J.~Malmstr{\"o}m, M.~Bodeg$\rm\mathring{a}$rd, L. Stolt, U.~Rau, H.~W.~Schock, and J.~H.~Werner, \href{https://doi.org/10.1016/S0040-6090(00)01726-0}{\textit{Thin Solid Films}}~\textbf{387}, 11  (2011).

\bibitem{Casper2016}
P. Casper, R. H\"{u}nig, G. Gomard, O. Kiowski, C. Reitz, U. Lemmer, M. Powalla, and M. Hetterich,  \href{https://doi.org/10.1002/pssr.201600018}{\textit{Phys. Status Solidi Rapid Res. Lett.}}~\textbf{10}, 376  (2016).	

\bibitem{Yang2016} 
K.-J.~Yang, D.-H.~Son, S.-J.~Sung, J.-H.~Sim, Y.-I~Kim, S.-N.~Park, D.-H.~Jeon, J.~Kim, D.-K.~Hwang, C.~W.~Jeon, D.~Nam, H.~Cheong, J.-K.~Kang, and D.-H.~Kim,  
\href{https://doi.org/10.1039/c6ta01558a}{\textit{J. Mater. Chem. A}}~\textbf{4}, 10151  (2016).
	
\bibitem{Gloeckler-Sites2005JPCS} 
M.~Gloeckler and J.~R.~Sites, \href{https://doi.org/10.1016/j.jpcs.2005.09.087}{\textit{J. Phys. Chem. Solids}}~\textbf{66}, 1891  (2005).	




\end{thebibliography}
\end{document}